\documentclass[3p]{elsarticle}

\usepackage{hyperref}
\usepackage{amsmath}
\usepackage{amssymb}
\usepackage{comment}
\usepackage{float}

\usepackage{subfigure}
\usepackage{tikz}
\usepackage{graphicx,xcolor}

\usepackage{multirow}

\usepackage[ruled,vlined,linesnumbered]{algorithm2e}

\newcommand{\compresslist}{
\setlength{\itemsep}{1pt}
\setlength{\parskip}{0pt}
\setlength{\parsep}{0pt}
}

\journal{Journal of \LaTeX\ Templates}

\biboptions{authoryear}

\usetikzlibrary{arrows,shapes,positioning,shadows,trees}

\tikzset{
  basic/.style  = {draw, text width=2cm, drop shadow, font=\sffamily, rectangle},
  root/.style   = {basic, rounded corners=2pt, thin, align=center,
                   fill=blue!20},
  level 2/.style = {basic, rounded corners=6pt, thin,align=center, fill=blue!20,
                   text width=8em},
  level 3/.style = {basic, thin, align=left, fill=pink!20, text width=7.5em}
}

\begin{document}

\begin{frontmatter}

\title{On Designing A Questionnaire Based Legacy-UI Honeyword Generation Approach For Achieving Flatness
 }

\author[iit]{Nilesh Chakraborty\corref{cor1}}
  \ead{nilesh.pcs13@iitp.ac.in}
\author[nit]{Shreya Singh}  
\author[iit]{Samrat Mondal}
  
  \cortext[cor1]{Corresponding author}
  \address{Department of Computer Science and Engineering}
  \address[iit]{Indian Institute of Technology, Patna}
  \address[nit]{National Institute of Technology Patna}
  \address{Bihar - 801103, India}

\begin{abstract}

Modern trend sees a lot usage of \textit{honeywords} (or fake password) for protecting the original passwords in the password file. However, the usage of \textit{honeywords} has strongly been criticized under the different security and usability parameters. Though many of these issues have been successfully resolved, research in this domain is still facing difficulties in \textit{achieving flatness} (or producing the equally probable \textit{honeywords} with reference to the original password). Though recent studies have made a significant effort to meet this criterion, we show that they either fall short or are based on some unrealistic assumptions. To practically fulfil this flatness criterion, we propose a questionnaire-oriented authentication system based on the episodic (or long term) memory of the users. Our study reveals that proposed mechanism is capable of generating significantly improved flatter list of \textit{honeywords} compared to the existing protocols. Subsequent discussion shows that the proposed system also overcomes all the limitations of the existing state of arts with no lesser than $95\%$ goodness.

\end{abstract}


\end{frontmatter}


\section{Introduction}
\label{sec:introduction}
All human beings have three lives: public, private, and secret - Gabriel Garcia Marquez.\\

Password based authentication remains as the most popular form of identity verification because of its high usability standard. However, different attack models have been launched over the time to diminish the security standard far below to the desired level. In recent days, password leakage from the compromised severs is booming gradually as the reports suggest that password breach at the server side heavily dominates the other security issues on the password. For example, some recent password data breaches include Adobe ($150$ million), Evernote ($50$ million), Anthem ($40$ million) \citep{eve-1}, RockYou ($32$ million) \citep{eve-3}, Tianya ($30$ million), Dodonew ($16$ million), 000webhost ($15$ million) \citep{eve-2}, Gmail ($4.9$ million) and Phpbb ($255$K) just to name a few. The security community has mainly identified two factors that causing password leakage from the different servers.
 
\begin{itemize}\compresslist
\item[1.] Human chosen passwords are highly skewed. Highly skewed passwords increase the predictability, building the success stories of the modern password cracking algorithms \citep{weir} \citep{ma}.
\item[2.] Recent password crackers utilize GPU-based password cracking technique which reduces time complexity of the password cracking to a great extent,  turning the things completely in favour of the attackers. For example, at Password'12, Gosney \textit{et al.} showed that $12$ hours is sufficient to brute force the entire key space of $95^8$ \citep{hpc-password-cracking}. This infers that with the help of GPU-based password cracking,  any password of length $8$ can be inverted from its hashed format within a time span of $12$ hours.
\end{itemize}

For addressing (or detecting to be precise) such serious threat, in $2013$, Juels and Rivest take one of the first significant initiatives to propose the fake password (or \textit{honeyword}) based framework \citep{honeyword-juels}.  Their work carries tremendous potential in handling such threat and because of this, modern days' research shows an inclination towards \textit{honeywords} in many password related domains \citep{honey-encrypt} \citep{honeyword-network}.\\

To be acceptable, like other security protocols, a \textit{honeyword} based authentication technique (or \textbf{HBAT}) must satisfy certain security and usability parameters. Before going detail into these parameters, we first present a brief overview of HBAT's working principle.

\subsection{Working principle of HBAT}
In HBAT, along with the original password of the users, the system stores some fake passwords (or \textit{honeywords}) in the password file. The index position of the original password is maintained in another file, belonging to a different server, known as \textit{honeyChecker}. The following presents an example of this kind of setup for the username and password as \textit{alice} and \textit{alice$2004$}, respectively.\\

\begin{center}
\begin{tabular}{c c c}
alice2002 & alice2006 & \textbf{alice2004}\\
alice1998 & alice2000 & alice2008
\end{tabular} 
\end{center}
In the above example, except \textit{alice$2004$}, the others are instances of \textit{honeyword}, generated by the system. The \textit{honeywords} and original password are collectively referred as \textbf{\textit{sweetwords}}. Along with the username (\textit{i.e.,} \textit{alice}) the index position of the original password (here $3$) is stored in the \textit{honeyChecker}.\\

Readers can notice that a compromised password file discloses all these \textit{sweetwords} to an adversary. However, it never reveals any information about the original password. Moreover, the adversary gets confused while identifying the original password of the user. On treating a \textit{honeyword} as the original password of the user, the index position of that \textit{honeyword} never finds a match in the \textit{honeyChecker}. In contrast, submitted password by a legitimate user always belongs to the index position which is same as the stored index value in the \textit{honeyChecker}.\\ 

If a mismatch occurs, the \textit{honeyChecker} directs a security alarm to the system administrator. On getting a match, the \textit{honeyChecker} sends a positive feedback to the system administrator, stating that the login request can be permitted.\\

 Therefore, a list of $k$ \textit{sweetwords} can sense the threat (\textit{i.e.,} password file has been compromised) with the probability $(k-1)/k$. It is \textit{important to note} that HBAT works under the distributed security framework which is assumed as much harder to compromise as a whole \citep{sauth} \citep{honeyword-juels}. Based on the presented text so far, next, we discuss the essential security and usability criteria that an HBAT must satisfy.

\subsection{HBAT security parameters}
\label{intro-2}
There are two well defined security parameters that an ideal HBAT must satisfy.
\begin{itemize}\compresslist
\item[(a)] Denial of services (DoS) resiliency. 
\item[(b)] Achieving flatness.
\end{itemize}
Initially, multiple system vulnerabilities (or MSV)  considered as another important security aspect \citep{honeyword-juels}. However, some of the salient research in this domain has treated this with much lesser significance \citep{flatness}. As the name suggests, in MSV, an attacker needs to compromise a distributed framework for threatening the security. Therefore this parameter violates the basic assumption of HBAT (\textit{i.e.,} distributed security is much harder to compromise as a whole). Hence, as advised in \citep{flatness}, we have only considered DoS resiliency and flatness as the essential security parameters.\\

\textbf{DoS resiliency:} Knowing the original password of a user (may be after performing the shoulder surfing attack \citep{weak-ssa-fc}), if an adversary can guess the stored \textit{honeywords} in the password file, then she may intentionally submit a \textit{honeyword}. The system then falsely interprets that the password file has been compromised. Under this situation, based on the normal HBAT security routine, the system denies any further login attempt from the users and blocks all the registered users' accounts. \textit{Chaffing} methods in \citep{honeyword-juels} are known to provide weak security against the DoS attack. For example, for the original password \textit{dextra}$5$ and $k = 6$, let the \textit{chaffing-by-tweaking-digit} 
in \citep{honeyword-juels} generates the following list of \textit{sweetwords}
\begin{center}
\begin{tabular}{c c c}
dextra$2$ & \textbf{dextra5} & dextra$1$\\
dextra$6$ & dextra$8$        & dextra$4$ 
\end{tabular}
\end{center}
Therefore, knowing the original password, \textit{dextra}$5$, an adversary can hit a \textit{honeyword} with the probability $5/9$. As suggested in \citep{kamouflage}, an ideal HBAT must take appropriate security measures to handle this kind of threat by generating \textit{hard-to-guess} \textit{honeywords}.\\ 

\textbf{Achieving flatness:} Ideally, all $k-1$ \textit{honeywords}, maintained against a user's account, should look like the original password of the user to obfuscate the attackers with the probability $1/k$. A list, containing $k$ equally probable \textit{sweetwords} is identified as the \textit{flat} list of \textit{sweetwords}.  However, there are many issues (\textit{e.g.}, a correlation between a username and password, targeted guessing \citep{honeyword-juels} \textit{etc.}) which increase an attacker's chances in identifying the original password of the users. Let the chosen username and password by a user are \textit{alex} and \textit{alex}$1992$, respectively. It is quite clear that the username and password in this example are correlated. Therefore, for $k =6$, if a system generates following list of \textit{sweetwords} based on the modelling-syntax-approach, proposed in \citep{kamouflage},
\begin{center}
\begin{tabular}{c c c}
jhon$2064$ & mery$1928$ & tias$7348$\\
cina$1990$ & baby$2368$ & \textbf{alex1992}
\end{tabular} 
\end{center}
then the adversary can easily identify the original password from the username \textit{alex}.\\

A list of \textit{sweetwords}, not meeting this criterion, violates the basic objective of an HBAT. Thus achieving flatness is absolutely important for an HBAT to hold its water. Though some serious efforts have been made to fulfill this prime security objective, most of them fall short either basis on hard to satisfiable assumptions or because of their design.\\

\subsection{HBAT usability parameters}
\label{intro-3}
Along with the aforementioned security parameters, there are following important usability specifications which play a significant role in determining the acceptability of an HBAT.       
\begin{itemize}\compresslist
\item[(a)] System interference.
\item[(b)] Stress on memorability.
\item[(c)] Typo-safety.
\end{itemize}

\textbf{System interference and stress on memorability:} These two usability specifications are complementary to each other. In their contribution in \citep{honeyword-juels}, authors show that for satisfying the previously mentioned security objectives, some HBAT may interfere in the password choice of the users and force them to remember an extra information for login. This additional information is used for generating the \textit{honeywords} for such HABT. Therefore, an HBAT, having system interference, puts stress on the users' mind. In a nutshell, user interface (UI) of such systems tells a user about the use of \textit{honeywords} and interact with the user to influence her password choice. These systems are categorized as \textbf{modified-UI} HBAT \citep{honeyword-nilesh}.\\

On the other hand, a \textbf{legacy-UI} HBAT does not tell the user about the use of \textit{honeywords}, nor interact with her to influence her password choice. Therefore this kind of HBAT does not have any system interference and hence puts no stress on user's mind. It is quite intuitive (and obvious) that a legacy-UI HBAT provides more user friendliness compared to a modified-UI HBAT and hence, an ideal HBAT should validate the users through the legacy-UI for providing a better usability standard \citep{flatness}.\\


\textbf{Typo-safety:} An HBAT would also like it to be rare for a legitimate user to set off a security alarm by accidentally entering a \textit{honeyword}. Typos are one possible cause of such accidents. Like in \citep{kamouflage}, there are HBAT which are capable of generating quite different \textit{sweetwords} from each other for becoming more typo-safe. In contrast, \textit{chaffing} methods in \citep{honeyword-juels} are considered as less typo-safe and thus threat the usability standard. For example, as discussed earlier, \textit{chaffing-by-tweaking-digit} may generate following list of \textit{sweetwords} for the password \textit{dextra}$5$ and $k = 6$.\\
\begin{center}
\begin{tabular}{c c c}
dextra$2$ & \textbf{dextra5} & dextra$1$\\
dextra$6$ & dextra$8$        & dextra$4$ 
\end{tabular}
\end{center}
Now, while entering the digit after entering the prefix, \textit{dextra}, successfully, chances of accidental submission of \textit{honeyword} will be $5/9$ . Thus \textit{chaffing-by-tweaking-digit} is not much typo-safe.

\subsection{Motivations and Contributions}
\label{intro-4} 
Based on the presented text so far, this section first builds the motivations behind this work and then lists the highlights of contributions.\\

\textbf{Motivation 1:} For achieving the desired security objectives, most of the recently proposed HBAT interfere in the password choice of the users. Therefore, the recent trend shows more inclination towards modified-UI HBAT rather than the legacy-UI. As discussed earlier, a modified-UI HBAT not only degrades the usability standard but sometimes it also ends up being unusable in practice (\textit{e.g., take-a-tail} in \citep{honeyword-juels}).  Recently, based on his proposed legacy-UI HBAT, \textit{Imran Erguler} has made a significant effort for satisfying the HBAT security parameters. However, one of our contributions shows that claimed security standard in \citep{flatness} is clearly overestimated and the method fails to achieve unconditional flatness. Thus there exists a gap in achieving flatness using a legacy-UI HABT. \\

\textbf{Motivation 2:} To be acceptable, assuring user friendliness is a must criterion and password is the highest form of authentication in providing that. Therefore, we also explore different authentication services that accept the passwords as input and put no stress on the users' mind. Our survey reveals that recognition is much easier than recall \citep{recognition-recall-01} \citep{recognition-recall-02}, and hence, recognition-based authentication systems are capable of providing very high usability standard. Introduced by Barton and Barton in $1984$ \citep{memoPass-1984}, questionnaire based authentication remains as one of the most popular recognition-based identity verification techniques till date \citep{memoPass-2013}. However, these questionnaire based authentication services acquire large memory space for storing the necessary information on per user basis. This is considered as a major overhead to be used as a practical solution.\\

\textbf{Motivation 3:} We also believe that if a questionnaire based authentication technique is proposed based on the memorable events of a user's life, then recalling/recognizing those events would not put any significant stress on that user.\\   

Mainly motivated by the aforementioned facts, below we present the highlights of our findings.\\ 


\textbf{Contribution 1:} We revisit the \textit{Imran Erguler's} approach and show that proposal in \citep{flatness} clearly overestimates the security standard for achieving the flatness. Also, our analysis reveals that to meet the claimed security standard, some of the assumptions made by the author degrades the usability standard substantially. \\

\textbf{Contribution 2:} We propose a questionnaire based authentication system based on the episodic (or long term) memory \citep{memoPass-2013} of the users. Our proposal not only reduces the storage cost significantly compared to conventional questionnaire based authentication systems but also meet the essential usability and security criteria (including achieving flatness) of an ideal HBAT.\\

\textbf{Contribution 3:} Finally, we have performed an exhaustive experimental analysis to show that proposed method also satisfactorily meet the conventional usability parameters (\textit{i.e.,} login time and error rate during login) and hence can be considered as a deployable solution in practice.\\

\textbf{Roadmap:} The rest of the paper is organized as follows. Section \ref{background} builds background of our work. Section \ref{dofl} discusses how to satisfy the important security parameters of HBAT in a best possible way. Section \ref{erguler-approach} finds vulnerabilities in a recently proposed \textit{honeyword} generation approach to show that there is a scope of improvements in this domain. Section \ref{prop} introduces the proposed mechanism. Section \ref{storage-ana} gives an overview of required storage to support our scheme. Following this, Section \ref{u-ana} and Section \ref{s-ana} detail about the usability and security standard of the proposed scheme, respectively. Section \ref{comparative-ana} compares our scheme with the most recent HBAT. Finally, Section \ref{conclusion} concludes the discussion. 

\section{Background} 
\label{background}
In spite of having other alternatives (\textit{e.g.,} graphical password), the users' community largely accepts textual password because of \textit{simple login interface} and \textit{adaptability}. However, unknowingly, the majority of the users put their password at risk \citep{differences-reality-think}, as the chosen passwords by them are highly skewed in nature \citep{skew-PIN} \citep{entropy2}. The unevenness in the password dataset plays a major role in the success stories of modern password cracking algorithms \citep{skew-password}. To detect the threat of password cracking, the \textit{honeywords} have successfully been used in many scenarios \citep{honeyword-network}, \citep{honey-encrypt}. However, most of the existing HABT, especially legacy-UI HBAT, fail to meet flatness criterion which remains a serious concern to the security community.

The questionnaire based authentication services have a tendency of using a user's specific information for forming the password. Though the influence of social networks (\textit{e.g.,} facebook, twitter) brings a portion of users' private lives in public, however, we believe that a careful selection of questionnaires still refrains the adversaries from knowing users' passwords (by exploring their network graph). In other words, we show that a carefully designed questionnaire based authentication system can truly mitigate the risk associated with targeted guessing attack \citep{targeted-guessing}. As targeted guessing is one of the major causes that restricts HBAT from achieving flatness, hence, a sense of eliminating this factor can be considered as a primary motivation behind this work. 

Attempts have been made since long to design an authentication system that accepts passwords derived from the users' memory \citep{memoPass-1990}. To the best of our knowledge, the contribution of Barton and Barton was among the very first few initiatives made in this direction \citep{memoPass-1984}. In their study, the representation of long term memory was classified into \textit{semantic} and \textit{episodic} and authors showed the superiority of latter one for forming the password. Lately, the proposal from Toshiyuki Masui explored the strength of the episodic memory in forming the passwords \citep{memoPass-2013}. Recent studies show some shift in utilizing the users' memory as the focus is getting more towards users' behavior or activity for generating the passwords \citep{activpass} \citep{activpass-compsec}. 

As mentioned before, the secrecy of users' private life has been changed drastically due to the interference of social network. Therefore, the past studies in this domain, like in \citep{memoPass-1990} \citep{memoPass-1984}, are losing their potential of becoming the valid alternatives for incorporating the HBAT. On the other hand, proposed methods in \citep{activpass} \citep{activpass-compsec} $-$ capable of providing good security against targeted guessing attack, depend on the fragile memory of the users. For example, the system in \citep{activpass-compsec} requires users to remember frequencies of their daily activities for authentication.  This type of login setups often increases the failure rate as some of these activities may not be that influential. Also, the questionnaires may be of dynamic nature \footnote{Before its appearance, a user cannot predict a dynamic question \citep{memoPass-2005}.}. 

Notably, in \citep{memoPass-2013}, authors' proposed system based on the episodic memory can be considered as the closest one to our contribution. However, their approach lags in different aspects. For example, it allows user specific questions which increase the memory overhead. Also, the selected questions may not always defend the targeted guessing attack. Moreover, during the registration phase, a user needs to provide the wrong alternatives for each question along with the correct answer. This may open up the door for the adversaries as the user may casually provide any random (or non-meaningful) information while submitting the wrong alternatives. Along with tackling all these existing issues, our proposal here incorporates HBAT for generating the flat list of \textit{sweetwords} to detect the password cracking.

\section{Criteria for DoS resiliency and achieving flatness}
\label{dofl}
As mentioned in Section \ref{intro-2}, there are two well defined security parameters associated with an HBAT $-$ (a) DoS resiliency and (b) achieving flatness. Earlier, we present a brief introduction to these two parameters. Therefore, in this section, we mainly focus on the necessary criteria that an HBAT must satisfy for handling these security issues. \\

\textbf{Essential criteria for handling DoS:} To be short, with the knowledge of an original password, an adversary can only mount DoS attack if she can guess any \textit{honeyword} with high probability. Therefore, a countermeasure will be reducing this chances of guessing.
Let \textit{n} be the number of all possible \textit{honeywords} that can be generated from the original password. Now, if a HBAT only selects $k-1 (k > 1)$ \textit{honeywords} to lure the attackers, then the probability of success of mounting the DoS attack becomes $(k-1)/(n-1)$. With the understanding of this, below we present the essential criterion for avoiding this attack. \\

\textbf{Criterion 1:} The value of \textit{n} should be significantly larger than $k$ so that $\dfrac{(k-1)}{(n-1)}$ yields to a very small value.\\

Literature in this domain shows that because of the large value of \textit{n}, modelling-syntax-approach in \citep{kamouflage} is DoS secure. In contrast, chaffing methods in \citep{honeyword-juels} provide weak security against DoS attack as values of \textit{n} and $k$ do not differ by much.\\

\textbf{Essential criteria for achieving flatness:} As discussed earlier, to be flat, all the \textit{sweetwords} must seem like equally probable to the adversaries. But there are a few factors which create a bar.\\

(a) \textit{Issue related to correlational hazards:} If there exists a correlation between the username and password, then an adversary can easily distinguish the original password from the list of \textit{sweetwords}. For example, if a user chooses her username and password as \textit{maradona} and \textit{football}, respectively, then existing legacy-UI HBAT provide no mechanism for masking the original password.\\

(b) \textit{Issue related to recognizable patterns in passwords:} Leaked password databases show that a large set of users use some particular patterns in their passwords which can be related to some well defined objects or facts. For example, \textit{bond007}, \textit{james007}, \textit{007007} are some instances of the passwords which were found to be used by many users as reported in \citep{10000-common-passwords}. The well known password patterns help an adversary in identifying the original password.\\

(c) \textit{Issue related to targeted password guessing:} Many users have a tendency of using their personal information for forming the passwords which can easily be gettable from their social network graphs. Therefore, it is often possible to deanonymize the users based on their social network graph \citep{targated-password-guessing-social-network} \citep{targated-password-guessing}. Also, given a user's identity, there are many ways to find biographical or demographic data about her online $-$ by exploiting information published on social networks, for example \citep{targeted-guessing}.

All existing HBAT (including the modified-UI also) fail to achieve flatness against this attack model, as the adversaries can distinguish the original password based on the information available in the social network.\\

\textbf{Criterion 2:} For achieving flatness, an HBAT must ensure that chosen login credentials by the users do not face any of the above issues.\\

Discussion in this segment helps the readers to understand the basic requirements for satisfying the HBAT security parameters. Based on the discussion in this section, we revisit the Erguler's approach in Section \ref{erguler-approach}.

\section{Revisiting Erguler's Approach}
\label{erguler-approach}
The core of the Ergular's method is to select $k-1$ \textit{honeywords} from the existing users' password in the password file. For example, system may assign following list of \textit{swettwords} against a user account for $k = 8$\\

\begin{center}
\begin{tabular}{c c c c}
happy8 & player & canadian & \textbf{87@lex}\\
memory & pound880 & love26 & 123456\\
\end{tabular}
\end{center}

where \textit{87@lex} is the original password of the user and rest are the passwords of other users that are selected as \textit{honeywords} for this particular user's account.\\

We believe that contribution of Erguler mainly aims to satisfy the basic usability criteria of an ideal HBAT. Notably, using this approach, a user does not perform any additional login activity or remember any extra information for generating the \textit{honeywords}. Also, chances of accidental submission of a \textit{honeyword} are very less as the \textit{sweetwords} are expected to be substantially different in structure. Therefore, this emerges as a pretty solid approach from the usability point of view. However, we have found that claimed security standard of the approach is questionable as it poorly copes with \textit{Criterion 1} and \textit{Criterion 2}. The identified major drawbacks of this method are listed below.

\begin{itemize}
\item \textbf{Drawback 1:} For satisfying the flatness criterion, a method must ensure that an adversary should not have any advantage in selecting the original password from the list of $k$ \textit{sweetwords}. However, Erguler's approach does not fit with this always.  For example, if the user selects username as \textit{alex}, then (because of the correlational hazardous)  an adversary can easily identify the original password (\textit{87@lex}) from the list of presented \textit{sweetwords}  in this section. Also, targeted guessing attack can greatly help an adversary in distinguishing the original password from the list of \textit{sweetwords}.  

\item \textbf{Drawback 2:} The proposal in \citep{flatness}, in Section $5.1$, shows that in a user population of size \textit{N}, if an adversary creates \textit{m}$+1$ users' account with a common password $p_{\text{Z}}$ for first \textit{m} accounts, then the password $p_{\text{Z}}$  has the following probability of appearing as a \textit{honeyword} for (\textit{m}$+1$)$^{\text{th}}$ account  

\begin{equation}
\text{Pr}(\textit{p}_{\text{Z}} \in W_{m+1}) = 1 - (\dfrac{N-m}{N})^{k}
\end{equation}
where $k$ and $W_{m+1}$ denote number of \textit{sweetwords} and list of \textit{sweetwords}, respectively, for the $(m+1)^{\text{th}}$ account.\\

This equation clearly suggests that the system selects \textit{honeywords} by following the \textit{with replacement strategy}. In other words, a system can select the same \textit{honeyword} multiple times. This leads towards the following major concern

\begin{itemize}
\item $\widehat{k}$ $(< k)$ repetition of a \textit{honeyword} in the list of $k$ \textit{sweetwords} increases the probability of successful password cracking to $\frac{1}{k-\widehat{k}}$ and this reduces the flatness.

\end{itemize}  
\item \textbf{Drawback 3:} Many systems (\textit{e.g.,} gmail) restrict the users to choose an existing username. Therefore, an adversary can get to know \textit{n} usernames without even compromising the password file.  As popular passwords are more likely to use as \textit{honeywords} because of their higher occurrence in the password file, thus this facilitates the following threat.\\

 Let us make a fair assumption that $30\%$ of the \textit{N} users select their password from the set of most popular passwords. Let the set be denoted as \textit{S}$_\text{pop}$. It would not be unfair to assume that because of the so many password leakages, the set \textit{S}$_\text{pop}$ is known to the adversaries. Therefore, the likelihood of none of them participates as a \textit{honeyword} for the $i^{th}$ user account can be presented in the form of the following equation

\begin{equation}
\text{Pr}(e(S_{\text{pop}}) \notin W_i) = (\dfrac{N-\frac{3N}{10}}{N})^{k-1}
\end{equation}
where \textit{e}($S_{\text{pop}}$) denotes an element from the set $S_{\text{pop}}$ and $W_i$ implies the list of \textit{sweetwords} for the $i^{th}$ user.\\

For a standard value of $k$ as $20$ \citep{honeyword-juels}, the above equation suggests that for a user account, there are only $0.11\%$ chances that none of the popular passwords from the set  $S_{\text{pop}}$ has been selected as a \textit{honeyword}. Therefore, even if the adversary knows less than $1\%$ of the usernames from a population of \textit{N} users, then also she can deterministically hit a \textit{honeyword} for those retrieved usernames.


The aforementioned discussion suggests that Erguler's method provides weak security against the DoS attack and poorly meets \textit{Criterion $1$}.

\item \textbf{Drawback 4:} For refraining the users from using the common passwords, Erguler proposed that during registration, a system may allow a password based on certain rules. For example, \textit{comprehensive8} enforces users to choose a password of length $8$, containing an uppercase and lowercase letter, a symbol, and a digit and not include a dictionary word \citep{password-guess-attack}. Also, under \textit{basic16} policy, the selected passwords by the users must hold at least $16$ characters in it \citep{password-guess-attack}. However, many users struggle to create and recall their passwords under strict password-composition policies  \citep{password-policy}. For example, reports suggest that within three months from password creation, $4.28\%$ of users often forget those passwords, chosen according to their own conveniences \citep{forgettablePassword}. Therefore, for maintaining its acceptability in a wide range, a system cannot always abide by these policies in practice.
\end{itemize}      

\textbf{Note 1:} The above analysis shows the limitation of Erguler's work. Also, meeting both the criteria (\textit{Criterion} $1$ and \textit{Criterion} $2$) together is still not being achieved satisfactorily by any of the existing HBAT. This further motivates us to propose an efficient solution that satisfies both the security objectives without violating the HBAT usability standards.

\section{Proposed methodology}
\label{prop}

\begin{figure}[!ht]
\begin{center}
\begin{tikzpicture}[
  level 1/.style={sibling distance=40mm},
  edge from parent/.style={->,draw},
  >=latex]

\node[root] {Questionnaire}
  child {node[level 2] (c1) {Question}}
  child {node[level 2] (c2) {Alternatives}};

\begin{scope}[every node/.style={level 3}]
\node [below of = c1, xshift=15pt] (c11) {Based on memorable event};
\node [below of = c11] (c12) {Generic in nature};
\node [below of = c12] (c13) {Based on shareable secret life};

\node [below of = c2, xshift=15pt] (c21) {Equally probable};
\node [below of = c21] (c22) {Generated from a finite set};

\end{scope}

\foreach \value in {1,2,3}
  \draw[->] (c1.191) |- (c1\value.west);

\foreach \value in {1,...,2}
  \draw[->] (c2.190) |- (c2\value.west);

\end{tikzpicture}
\end{center}
\caption{Skeleton of our proposal for achieving flatness.}
\label{overview}
\end{figure}
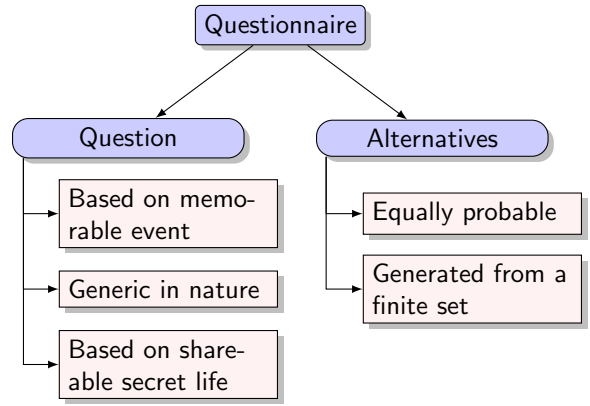

Presented text so far suggests that the amalgamation of conventional password based authentication techniques and HBAT has failed to achieve flatness. Therefore, in this study, we explore questionnaire based authentication technique as we believe it can satisfy few essential requirements for achieving the desired objective. Also, questionnaire based authentication allows a user to recognize the correct alternative which is much easier than recall \citep{recognition-recall-01}. Figure \ref{overview} shows the basic properties of proposed questionnaire based authentication service. Rest of the section unfolds the following two major topics.
\begin{itemize}
\item How our contribution meets these properties.
\item How flatness can be achieved by satisfying these properties. 
\end{itemize}

As the above figure suggests, each question must have the following $3$ properties. 
\begin{itemize}
\item \textbf{Property 1:} Answer of each question must be related to some memorable events. 
\begin{itemize}
\item \textbf{Objective 1:}This property ensures that recalling the correct alternative does not put any stress on the users' mind.
\end{itemize}

\item \textbf{Property 2:} The questions must be generic in nature.
\begin{itemize}
\item \textbf{Objective 2:} Most of the users can relate themselves to a generic question. Therefore, a generic question set can be used to maintain the same set of question for all the users. This has a huge impact in minimizing the storage cost (ref. to Section \ref{storage-ana}).
\end{itemize} 
\item \textbf{Property 3:} The questions must belong to the shareable secret lives of the user.
\begin{itemize}
\item \textbf{Objective 3:} This property ensures that the information related to the questions cannot be obtained from the social network graph of the users. To be short, our system tries to meet this property for mitigating the risk associated with the targeted guessing attack. 
\end{itemize}
\end{itemize}

A person experiences lots of incidents in her life. Some of these incidents make a lasting impact on that person's mind (\textit{e.g.,} Name of the favourite teacher). Human psychology states that this unforgettable information (or events) belongs to the episodic memory of the users. For designing our system,  we use this kind of information to meet the \textit{property} $1$. 

For satisfying \textit{property} $2$, we propose a set of $Q$ generic questions, out of which the users need to choose and give answers for $q$ $(\leq Q)$ questions. As the users may sometimes feel uncomfortable in sharing their too personal information (even with the systems), therefore, we ensure that the selected $Q$ questions must belong to the \textit{shareable secret life} of the users and thus satisfying \textit{property} $3$.

\subsection{Survey for selecting the questions}
\label{prop-1}
Before modelling our questionnaire based authentication system, we first formed a small committee of $30$ people between an age group $24$ to $63$. All of them were academician and some of them ($17$ people) were from the human psychology background. The committee identified the events that are generic and belong to the episodic memory of the users. The committee members also suggested that the events related to the episodic memory can be of $3$ types.

\begin{itemize}\compresslist
\item \textit{Non shareable events:} An event that cannot be shared with either a system or person. For example, the event related to any self-committed crime.
\item \textit{Partially shareable events:} An event that cannot be shared with a person, but can be with a system. For example, \textit{Name of the person you hate secretly.}
\item \textit{Shareable events:} An event that can be shared with both the system and person. For example, \textit{Name of your best friend in childhood}?
\end{itemize} 
  
Finally, with the consent of all $30$ committee members, we select $10$ most generic questions based on the \textit{shareable secret events}. As we manually examine the genericness of the questions by asking them to the users, thus, we do not consider \textit{non shareable} and \textit{partially shareable} events. After selecting $Q = 10$ questions, we perform following two tasks.

\begin{itemize}\compresslist
\item \textbf{\textit{Task $1$:}} We explore two popular social networks (\textit{facebook} and/or \textit{twitter}) of $553$ users to search the answers for the selected questions. Unsurprisingly, we find that only $2.71\%$ users reveal the answers for at most $2$ questions out of the $10$ selected questions/questionnaires by us.

\item \textbf{\textit{Task $2$:}} We ask $179$ people (excluding the committee members) to give answers to no lesser than $6$ questions. For the questions/questionnaires presented below, Table \ref{mem-initial} shows the genericness of the selected questions.     
\end{itemize}


\begin{center} 
\noindent\fbox{%
    \parbox{0.98\textwidth}{ 
    \textbf{List of Questions}     
\begin{itemize}\compresslist
\item \colorbox{lightgray}{\hbox to 0.9\textwidth{$\mathbb{Q}1:$ What is name of your first childhood friend?\hfill}}
\item \colorbox{lightgray}{\hbox to 0.9\textwidth{$\mathbb{Q}2:$ What is the name of the movie you first saw in a cinema hall?\hfill}}
\item \colorbox{lightgray}{\hbox to 0.9\textwidth{$\mathbb{Q}3:$ What is the name of the doctor you visited often in childhood?\hfill}}
\item \colorbox{lightgray}{\hbox to 0.9\textwidth{$\mathbb{Q}4:$ What is the name of your parents friend you find (or found) your close too?\hfill}}
\item \colorbox{lightgray}{\hbox to 0.9\textwidth{$\mathbb{Q}5:$ What was your (or your closest one's) birth time?\hfill}}
\begin{itemize}
\item \colorbox{lightgray}{\hbox to 0.853\textwidth{(a) Morning			(b) Afternoon		(c) Evening		(d) Night.\hfill}} 
\end{itemize}
\item \colorbox{lightgray}{\hbox to 0.9\textwidth{$\mathbb{Q}6:$ What is the last two digits of the phone number you call often?\hfill}}
\item \colorbox{lightgray}{\hbox to 0.9\textwidth{$\mathbb{Q}7:$ What is the name of the person (except your parents) who gave you a special gift in childhood?\hfill}}
\item \colorbox{lightgray}{\hbox to 0.9\textwidth{$\mathbb{Q}8:$ Who is your favourite teacher?\hfill}}
\item \colorbox{lightgray}{\hbox to 0.9\textwidth{$\mathbb{Q}9:$ What was your best rank (or roll number) in school?\hfill}}
\item \colorbox{lightgray}{\hbox to 0.9\textwidth{$\mathbb{Q}10:$ Marriage anniversary of your parents (or closest one) falls in which quarter of the year?\hfill}}
\begin{itemize}
\item \colorbox{lightgray}{\hbox to 0.853\textwidth{(a) Jan-Mar			(b) Apr-Jun			(c) Jul-Sep		(d) Oct-Dec.\hfill}}
\end{itemize} 
\end{itemize}}}
\end{center}

\begin{table}[!h]
\centering
\begin{tabular}{|c|c|c|}
\hline
No. of questions & Successfully answered ($\%$) & Unable to answer ($\%$)\\
\hline
$6$  			 &			$100$					&	$0$\\
\hline
$7$				 &			$100$					&   $0$\\
\hline
$8$				 &			$97.7$					&   $3.3$\\
\hline
$9$				 &			$83.8$					&   $16.2$\\
\hline
$10$			 &			$67.6$					&   $32.4$\\
\hline
\end{tabular}
\caption{Memorability study based on the selected questions/questionnaires.}
\label{mem-initial}
\end{table}
   
The outcome of \textit{task} $1$ reflects that based on the information, available in the social network graphs of the users, it is hard to guess the answers of the selected questions. Also, the experimental study presented in Table \ref{mem-initial} reflects that selected questions are truly generic and based on the episodic memory of the users.\\ 

\textbf{Note 2:} In a nutshell, our survey shows that selected questions by us can mitigate chances of targeted guessing attack.  It is quite intuitive that answer of these questions has very little chance of having any correlation with the usernames. Also, the answers here do not follow any well known pattern. Therefore, the issues related to achieving flatness will not arise in this scheme. However, for satisfying flatness criterion completely, questionnaire based authentication system must generate a flat list of alternatives for each question. In Section \ref{prop-2}, we have tried to achieve this.

\subsection{Mechanism for generating the alternatives}
\label{prop-2}
Figure \ref{overview} (at the beginning of Section \ref{prop}) indicates that the alternatives must hold the following properties. 
\begin{itemize}
\item \textbf{Property 4:} All the alternatives corresponding to a questionnaire must be equally probable from the adversaries point of view.
\begin{itemize}
\item \textbf{Objective 4:} A set of equally probable alternatives makes it harder for the adversaries to choose the correct alternative.
\end{itemize}
\item \textbf{Property 5:} All the alternatives corresponding to a questionnaire must belong to a finite set.
\begin{itemize}
\item \textbf{Objective 5:} It can be noticed that a set of infinite elements (especially a set of nouns) may contain both meaningful and non-meaningful information. Therefore, if the alternatives are chosen from such an uneven set, then it would affect the flatness. On the other hand, it is really hard to control the nature of the elements belonging to an infinite set.  However, some uniformity may be imposed on a finite set. It is quite obvious that selected alternatives from a uniform set helps in achieving flatness. Therefore, for controlling the nature of the set, we restrict the cardinality of the set to a finite limit. Notably, this also has a huge impact in minimizing the storage cost (ref. to Section \ref{storage-ana}). 
\end{itemize} 
\end{itemize}

It can be noticed that $\mathbb{Q}5$ and $\mathbb{Q}10$ have predefined alternatives and the alternatives are absolutely flat. Also, alternatives corresponding to $\mathbb{Q}6$ belong to a finite range (from $00$ to $99$) and leak no clue about the correct option. In contrast, rest of the questions do not have any predefined alternatives and, their answer belongs to a set of infinite nouns. We identify these questions as \textbf{\textit{b}$\_$\textit{questions}}. For such a question, based on the answer of a user, 
\begin{itemize}\compresslist
\item \textbf{Strategy I:} either the system accepts the false alternatives from the user.
\item \textbf{Strategy II:} or the system may take the responsibility of generating the false alternatives.
\end{itemize} 
However, generated alternatives by following the \textit{strategy I} may not always produce flat alternatives. For example, for the question $\mathbb{Q}2$ ``\textit{what is the name of the movie you first saw in a cinema hall?}", the user may provide the correct answer and the false answer as \textit{Titanic} and \textit{Spotlight}, respectively. Therefore, the system may present the questionnaire in the following form during the login period.

\begin{itemize}
\item \colorbox{lightgray}{\hbox to 0.9\textwidth{What is the name of the movie you first saw in a cinema hall?\hfill}}
\begin{itemize}
\item \colorbox{lightgray}{\hbox to 0.853\textwidth{(a) Titanic			(b) Spotlight \hfill}} 
\end{itemize}
\end{itemize}   
To identify the correct alternative, an attacker first searches the release date of these movies. \textit{Titanic} and \textit{Spotlight} released in $1997$ and $2016$, respectively. Therefore, based on the user's age (or otherwise) the attacker may conclude that \textit{Titanic} is the more probable answer.

As discussed earlier, while entering the false alternatives, a user may reluctantly enter any random string. This also helps the adversary in identifying the correct answer.

Moreover, a human chosen answer and a system generated answer differ in structure. Therefore, \textit{strategy II} also fails to generate equally probable alternatives. In a nutshell, both the strategies fail to achieve the desired objective. \\

It can be noticed that a \textit{b}$\_$\textit{question} accepts an answer in the form of an alphabet string. Even if the answer contains digit in it (for $\mathbb{Q}2$ only) the digit can be converted to an alphabet string (\textit{e.g.,} $10$ is replaced by \textit{ten}). For generating flat alternatives for each \textit{b}$\_$\textit{question}, we have adopted the following strategy.

\begin{itemize}\compresslist
\item [1.] Instead of considering the whole answer, the system considers an alphabet from a particular index position. Let the index be denoted as $i$ ($1 \leq i \leq$ \textit{length of the answer}). The value of $i$ may vary depending on a question and $i'$s value is determined based on an experimental analysis, presented at the end of this section. 

\item [2.] System fix this alphabet as one of the \textit{d} alternatives.

\item [3.] Another \textit{d}$-1$ false alphabets are chosen from a \textbf{\textit{group of equally probable alphabets}} (an analysis on this is presented in this section).

\item [4.] With the question number, the system stores this \textit{d} length tuple in the database.

\item [5.]At first, the system identifies the appropriate question with the help of the question number, stored in the database. Thereafter, by treating each element of the corresponding tuple as an alternative, the system displays the appropriate questionnaire during the login period.
 
\item [6.] The user needs to recognize the $i^{th}$ character of her answer from the set of alternatives.     
 
\end{itemize} 

For example, for $\mathbb{Q}2$, if the user gives her answer as \textit{Titanic} now, then for the value of $i$ as \textit{length of the answer} (or \textit{last index}) and \textit{d} $= 4$, the system presents the questionnaire in the following form.

\begin{itemize}
\item \colorbox{lightgray}{\hbox to 0.9\textwidth{What is the name of the movie you first saw in a cinema hall? Recognize the last alphabet.\hfill}}
\begin{itemize}
\item \colorbox{lightgray}{\hbox to 0.853\textwidth{(a) V   \hspace{0.8cm}			(b) F  \hspace{0.8cm}     (c) C \hspace{0.8cm}  (d) Z \hfill}} 
\end{itemize}
\end{itemize}
System stores the information related to this questionnaire as $\mathbb{Q}2$, $<$V, F, C, Z$>$. Readers can notice that if the alternatives are equally probable, then it becomes hard for the adversary to identify the correct answer. Hence, it eliminates the issues related to \textit{strategy I} and \textit{strategy II}. It is \textbf{very important to note} that the alternatives are being represented in the form of alphabets. Therefore, we state that all the alternatives belong to the finite set of $26$ alphabets. Hence the \textit{property} $5$ is satisfied. Next, we have taken the initiatives to satisfy \textit{property} $4$.\\

\textbf{Generating flat alternatives:}  Previously, we mentioned that for each \textit{b$\_$question}, \textit{d} ($> 1$) alternatives will be chosen from a \textit{group of equally probable alphabets}. This also infers that there would not be much variation among the elements belonging to the same group. However, variation among the elements belonging to the different groups must be substantially high.\\

\begin{figure}[!ht]
\centering
\includegraphics[scale=1.4]{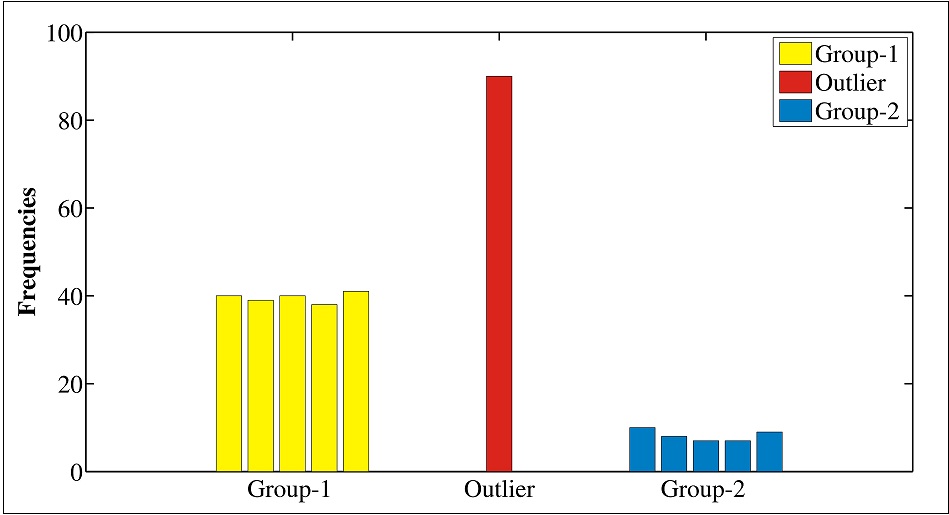}
\caption{Above figure shows that bars with the same color come under the same group because of similar frequencies. Outliers are just the odd elements that cannot be fitted in any group.}
\label{exm-grp}
\end{figure}

Figure \ref{exm-grp} indicates this property of the group where groups are formed based on the frequencies of the elements. The outlier shown in this figure indicates an element that falls beyond the distribution of the groups (\textit{e.g.,} an element with very high frequency compared to the other elements) and hence, cannot be incorporated in any group.\\

In this study, the groups have been formed based on the frequencies of the alphabets concerning to a particular index position of the answers. Literature in the field of human psychology shows that human can quickly recall first $3$ and the last letter of any alphabet string \citep{vstm}. Therefore, at first, we vary the index value among $1$, $2$, $3$ and \textit{last}. For a given \textit{dataset of answers}, then we measure the variation of alphabets for these index values. We only select that index value which gives a nice variation of alphabets, satisfying the group property as depicted in Figure \ref{exm-grp}. A minimum number of outliers also work in favour of selecting an index value.\\

Readers can notice that except $\mathbb{Q}1$, the answer of each \textit{b$\_$question} ($\mathbb{Q}2$, $\mathbb{Q}3$, $\mathbb{Q}4$, $\mathbb{Q}7$ and $\mathbb{Q}8$) is related to the name of a person. As the name of a person is highly influenced by the demographical information, therefore, in our study, we only consider the names of our country people, \textit{i.e.,} \textit{Indian} names. However, the answer of $\mathbb{Q}1$ may not always represent an \textit{Indian} movie as the foreign movies are also very popular in \textit{India}. \\

For capturing patterns in the \textit{Indian} names and movie names, we take reference of two datasets. For building the \textit{Population name} dataset (related to \textit{Indian} names), we collect almost $24000$ sample \textit{Indian} names (both male and female) from \citep{name-indian-people}. On the other hand, for forming the reference dataset for the questionnaire $\mathbb{Q}1$,  we collect $280$ most popular \textit{Indian} movie names in last $60$ years and the name of $190$ foreign movies that became popular in \textit{India}. We name this dataset as \textit{Movie name}. \\

\begin{figure*}[!ht]
\centering
\includegraphics[width=\textwidth]{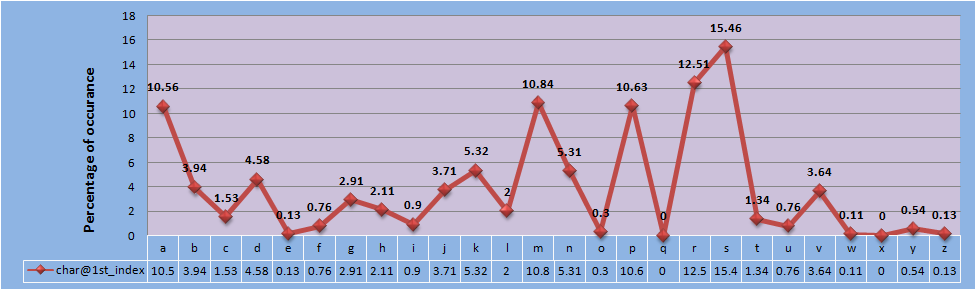}
\caption{Above figure shows frequencies of the alphabets for the \textit{Population name} dataset at index position $1$. The figure suggests that most of the \textit{Indian} names start with the alphabet \textit{``S"}.} 
\label{freq-1}
\end{figure*}
    
\begin{figure*}[!ht]
\centering
\includegraphics[width=\textwidth]{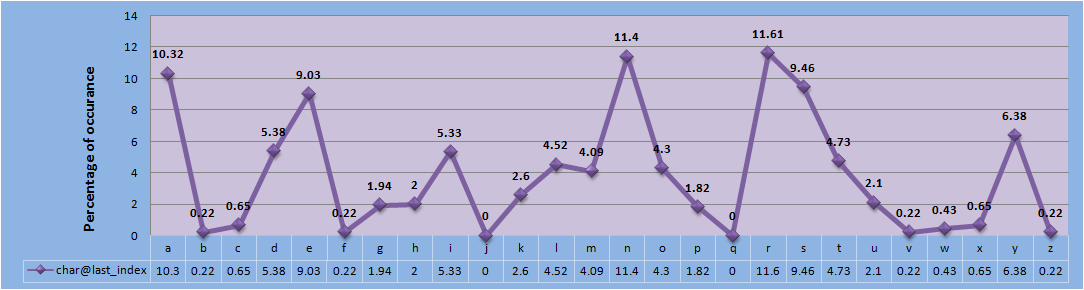}
\caption{Above figure shows frequencies of the alphabets for the \textit{Movie name} dataset at index position \textit{last}. The figure suggests that most of the movie names end with the alphabet \textit{``R"}.} 
\label{freq-2}
\end{figure*}

Note that we have made the frequency analysis of alphabets for all the previously mentioned index values for both these \textit{datasets}. However, figures for all these cannot be incorporated only because of space constraints. For two different datasets, Figure \ref{freq-1} (related to \textit{Population name}) and Figure \ref{freq-2} (related to \textit{Movie name}) here show the frequencies of alphabets for index position $1$ and \textit{last}, respectively. We choose these index values as they best satisfy the group formation criteria, depicted in Figure \ref{exm-grp}. \\

Above figures show no existence of outliers which also  motivates us for selecting these index values (\textit{i.e.,} $1$ and \textit{last} for \textit{Population name} and \textit{Movie name} datasets, respectively) over the others.  Based on the obtained frequencies of alphabets shown in Figure \ref{freq-1} and \ref{freq-2}, next, we focus on forming the groups. The groups here are formed based on the Algorithm \ref{algo-1}. The role of the each parameter used in this algorithm is described below.\\

\begin{algorithm}[H]
\SetAlgoLined
 initialization: curr$\_$freq $\leftarrow$ $\infty$, i $\leftarrow 1$\;
 \While{$(1)$}{
  \textit{peak} $\leftarrow$ \textit{nextHfreq}(curr$\_$freq, $\mathcal{D}$);\//* returns the immediate lesser value than curr$\_$freq from  $\mathcal{D}$ */\\
  \textit{base} $\leftarrow$ $\alpha \%$ of \textit{peak}\;
  \If{(\textit{base} $-$ $\epsilon_B \leq 0$)}{
  \textit{base} $\leftarrow$ $0$\;
  }  
   $\mathcal{G}_i \leftarrow$ \textit{gFr}(\textit{peak}, \textit{base}, $\mathcal{D}$);\//* returns those elements from $\mathcal{D}$, frequency of which fall between \textit{peak} and \textit{base} */\\ 
  \eIf{(\textit{base} $:= 0$)}{
   \textbf{break}\;
   }{
   curr$\_$freq $\leftarrow \beta \%$ of \textit{base} $+$ $\epsilon_P$\; i++ \; 
  }
 }
 \caption{\textbf{Forming$\_$Group}($\alpha$, $\beta$, $\epsilon_P$, $\epsilon_B$, $\mathcal{D}$)}
 \label{algo-1}
\end{algorithm}

\begin{itemize}\compresslist
\item $\mathcal{D}$ represents a dataset comprising of alphabets along with their frequency. In this study, $\mathcal{D}$ have been formed with references to Figure \ref{freq-1} and \ref{freq-2}.
\item $\alpha $ determines the lowest frequency in the group with respect to the highest frequency (\textit{i.e.,} \textit{peak}).
\item $\beta$  determines a minimum frequency gap between two consecutive groups.
\item $\epsilon_P$  is a very small value ($< 1$) which is used for slightly increasing the \textit{peak} for including the elements with closer frequencies.
\item $\epsilon_B$ is also a very small value ($< 1$) which is used in the algorithm for decreasing the \textit{base} value to $0$. This parameter come into play while \textit{base} value is very close to $0$. 
\end{itemize}

The groups returned by Algorithm \ref{algo-1} for the datasets $-$ pictorially represented in Figure \ref{freq-1} and \ref{freq-2}, are shown in Table \ref{groupTable}. It is \textbf{important to note} here that proposed system requires \textit{d} alternatives for each questionnaire. As the alternatives are selected from a group only, therefore, the size of each group must not be lesser than \textit{d}. We consider the default value of \textit{d} as $4$ here. Also, we state that based on the geographical location, the selected index values may differ, however, the questionnaire may remain unchanged as they are truly generic in nature.\\ 

\begin{table}[!h]
\centering
\resizebox{0.8\textwidth}{!}{
\begin{tabular}{c|c|c|c|c|c|l}
\cline{2-6}
&g$\_$id & $\mathbb{E}$ & elements & mean & variance \\ 
\cline{1-6}
\multicolumn{1}{ |c  }{\multirow{4}{*}{Population} } &
\multicolumn{1}{ |c| }{1} & \textbf{5} & $\{$\textbf{A, M, P, R, S}$\}$ & \textbf{11.9} & \textbf{3.49} &     \\ \cline{2-6}
\multicolumn{1}{ |c  }{}                        &
\multicolumn{1}{ |c| }{2} & \textbf{7} & $\{$\textbf{B, D, G, J, K, N, V}$\}$ & \textbf{4.2} & \textbf{0.7} &     \\ \cline{2-6}
\multicolumn{1}{ |c  }{}                        &
\multicolumn{1}{ |c| }{3} & \textbf{5} & $\{$\textbf{C, H, I, L, T}$\}$ & \textbf{1.57} & \textbf{0.19} &     \\ \cline{2-6}
\multicolumn{1}{ |c  }{}                        &
\multicolumn{1}{ |c| }{4} & \textbf{9} & $\{$\textbf{E, F, O, Q, U, W, X, Y, Z}$\}$ & \textbf{0.34} & \textbf{0.08} &     \\ \cline{1-6}\\
\cline{1-6}
\multicolumn{1}{ |c  }{\multirow{4}{*}{Movie} } &
\multicolumn{1}{ |c| }{4} & \textbf{5} & $\{$\textbf{A, E, N, R, S}$\}$ & \textbf{10.3} & \textbf{1.03}  \\ \cline{2-6}
\multicolumn{1}{ |c  }{}                        &
\multicolumn{1}{ |c| }{5} & \textbf{7} & $\{$\textbf{D, I, L, M, O, T, Y}$\}$ & \textbf{4.96} & \textbf{0.53}  \\ \cline{2-6}
\multicolumn{1}{ |c  }{}                        &
\multicolumn{1}{ |c| }{6} & \textbf{5} & $\{$\textbf{G, H, K, P, U}$\}$ & \textbf{2.09} & \textbf{0.07} &     \\ \cline{2-6}
\multicolumn{1}{ |c  }{}                        &
\multicolumn{1}{ |c| }{7} & \textbf{9} & $\{$\textbf{B, C, F, J, Q, V, W, X, Z}$\}$ & \textbf{0.29} & \textbf{0.05} &     \\
\cline{1-6}
\end{tabular}}
\vspace{0.15cm}
\caption{Groups returned by Algorithm $1$ with reference to the datasets shown in Figure \ref{freq-1} and \ref{freq-2}. $\mathbb{E}$ represents number of elements in a group. We set the values for $\alpha$, $\beta$, $\epsilon_P$ and $\epsilon_B$ as $45$, $85$, $0.1$ and $0.6$, respectively, for the dataset shown in Figure \ref{freq-1}. For the dataset shown in Figure \ref{freq-2}, we set the values for $\alpha$, $\beta$, $\epsilon_P$ and $\epsilon_B$ as $65$, $85$, $0.1$ and $0.6$, respectively.}
\label{groupTable}
\end{table}

After retrieving the alphabet from the submitted answer by a user, the system first identifies the group to which the alphabet belongs to. For the \textit{d}'s value as $4$, the system randomly chooses $4$ alphabets from the identified group, including the retrieved one. Along with the questionnaire number, the system saves these alphabets as a tuple in the database. During login, along with the questionnaire, the system then displays each element of the tuple as an alternative.\\

\textbf{Note 3:} Alternatives for each \textit{b$\_$question} are chosen from a group of almost equally probable elements. Thus we may claim that the alternatives will be almost equally probable from the attackers' point of view. Previously, it has already been discussed that except \textit{b$\_$questions}, others are capable of generating absolutely flat alternatives. Therefore, our proposal successfully holds \textit{property} $4$.   Also, \textit{property} $5$ has been satisfied early in the Section \ref{prop-2} to meet all the characteristics of the alternatives, depicted in Figure \ref{overview}. Moreover, Section \ref{prop-1} shows that our system also fulfils all the properties of questions, shown in Figure \ref{overview}. Therefore, presented text in Section \ref{prop} suggests that our proposal highly satisfies all the criteria for achieving flatness.

\subsection{Registration and login procedures}
\label{prop-3}
In the registration phase, a user needs to answer to $q$ questions out of $Q$ (here $10$) questions. The value of $q$ is predetermined and remains same for all the users. We suggest to vary $q$ between $6$ to $8$ as Table \ref{mem-initial} shows that users perform exceedingly well to answer to this many questions. Based on the answers submitted by the user, the system forms the tuples and saves them in the database. The user recognizes the correct answers corresponding to the $q$ questionnaires during login.\\

The users may adopt following two different login strategies for authentication purpose.
\begin{itemize}\compresslist
\item[1.] Naive login approach
\item[2.] Faster login strategy.
\end{itemize}

\textbf{Naive login approach:} In naive login approach, a user follows the conventional login steps for authentication purpose. In this process, the user first looks at a questionnaire and recognizes the correct answer. For submitting her response, the user then enters the option, referring to the correct answer. For example, if the user finds that \textit{option B} is denoting the correct answer, then she submits \textit{B} to the system. Therefore, by following this naive strategy, the user reads each question before submitting her choice for it.\\

\begin{figure}[!ht]
\centering
\includegraphics[width=0.7\textwidth]{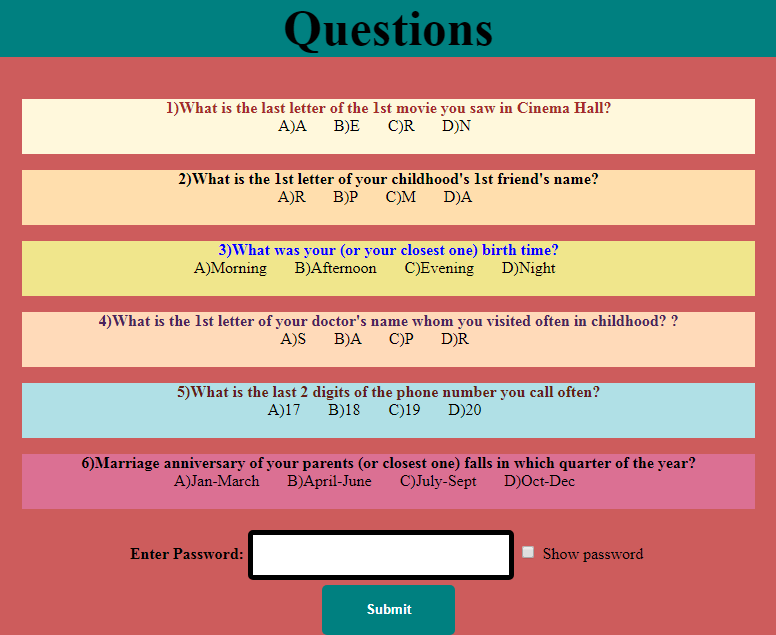}
\caption{An instance of the login interface for $6$ questionnaires for the proposed system.} 
\label{interface}
\end{figure}

\textbf{Faster login strategy:} With time, a user can develop a smarter login strategy for improving her login time. After using our system for a certain time,  $t_\text{mem}$, a user may able to remember the option sequence. For example, for giving answer to $6$ questionnaires, if the user selects the options ``\textit{C}", ``\textit{C}", ``\textit{B}", ``\textit{A}", ``\textit{B}" and ``\textit{D}", then the option sequence will be \textit{CCBABD}. Now, even without looking at the questionnaires, if the user submits this option sequence, then also the system will permit the login. In Section \ref{u-ana}, we have shown that the value of $t_\text{mem}$ is sufficiently less. For $q = 6$, Figure \ref{interface} shows an instance of the login interface. 
 
\section{Analyzing the storage cost}
\label{storage-ana}
In this section, we show that proposed strategy significantly reduces the storage cost compared to the existing questionnaire based authentication techniques. Our study here reveals that from the saved storage space, proposed system can successfully compensate the additional storage, required for incorporating the \textit{honeywords}. We denote (QBA)$_\text{SC}$ and (PQBA)$_\text{SC}$ as the required storage space for maintaining the login credentials in a conventional questionnaire based authentication technique and proposed questionnaire based authentication technique, respectively. Let S$_\text{SC}$ be the required memory space for maintaining a string.\\ 

Along with the \textit {d} alternatives, the conventional questionnaire based authentication techniques maintain $q$ (user specific) questions for each user. As both the question and alternative can be treated as a string, therefore, (QBA)$_\text{SC}$ can be represented in the form of the following equation.\\

\begin{equation}
\text{(QBA)}{_\text{SC}} = q \times \text{S}_\text{SC} + q \times d \times \text{S}_\text{SC} + 2 \times \text{S}_\text{SC}
\label{eqn-1} 
\end{equation} 
The term, $2 \times$ S$_\text{C}$, in Equation \ref{eqn-1} denotes the required memory space for storing the valid option sequence and the username. For the default value of \textit{d} as $4$, the above equation yields to  ($5q + 2$) S$_\text{SC}$.\\

Like in \citep{flatness}, proposed method maintains the login credentials in two different files in a server. We denote these two password files as $F_1$ and $F_2$. The information stored in $F_1$ helps in creating the login page for a user. 
Along with the username, $F_1$ stores all the questionnaire numbers answered by the user. The tuples for all questionnaires are treated as a single string. For the login page shown in Figure \ref{interface}, Table \ref{q-org} shows the contents of $F_1$.\\

\begin{table}
\begin{tabular}{|c| c| c|}
\hline
username & questionnaire number & tuple-string\\
\hline
alex	 & $\mathbb{Q}$2, $\mathbb{Q}$1, $\mathbb{Q}$5,    & $<$A, E, R, N$>$ $<$R, P, M, A$>$ $<$Morning, Afternoon, Evening, Night$>$\\
         & $\mathbb{Q}$3, $\mathbb{Q}$6, $\mathbb{Q}$10                     & $<$18, 19, 20, 21$>$ $<$Jan-mar, Apr-Jun, Jul-Sep, Oct-Dec$>$\\
\hline 
\end{tabular}
\caption{Above table shows an entry in $F_1$ for the username alex.}
\label{q-org}
\end{table}    

In $F_2$, together with username, the system stores $k$ option sequences, including the original one. Therefore, each option sequence can be treated as a \textit{sweetword} here. For $k =6$ and the correct option sequence as \textit{BDBAAA}, below we show an instance of $F_2$'s content.

\begin{center}
\begin{tabular}{c c c}
ADBCAB & \textbf{BDBAAA} & BAADAC\\
DADACA &     CCADAA	     & ABBBAA\\		
\end{tabular}
\end{center}
As discussed earlier, following the normal HBAT routine, the index position of the original \textit{sweetword} (here $2$) is stored in the \textit{honeyChecker} server. The information stored in $F_2$ incurs the normal storage overhead of HBAT and hence, (PQBA)$_\text{SC}$ is calculated based on the contents of $F_1$ only. This is shown in Equation \ref{eqn-2}.\\

\begin{equation}
\text{(PQBA)}{_\text{SC}} =  \text{S}_\text{SC} +  \text{S}_\text{SC} + \text{S}_\text{SC}
\label{eqn-2} 
\end{equation} 

For any particular value of $q$, the proposed protocol reduces the storage overhead and this can be represented in the form of the following equation.\\ 

\begin{equation}
\text{Saved storage space =} \dfrac{\text{(QBA)}{_\text{SC}}-\text{(PQBA)}{_\text{SC}}}{\text{(QBA)}{_\text{SC}}} \times 100\%
\end{equation}

Above equation yields to $90.62\%$, $91.85\%$ and $92.68\%$ for the chosen values of $q$ as $6, 7$ and $8$, respectively.\\

The information in $F_2$ requires  $(k  + 1) \times$ S$_\text{SC}$ memory space for storing each user's information. Therefore, for default value of $k$ as $20$, as suggested by authors in \citep{honeyword-juels}, the required memory space would be $21$S$_\text{SC}$. Notably, the required space for storing $10$ questions and the information shown in Table \ref{groupTable} do not depend on the population size and hence, for a large population of $\mathbb{N} = 100000$, the required space for storing this information can be treated as negligible. \\

\textbf{Note 4:} The discussion in this section shows that for considered value of $q$ as $6, 7$ and $8$, the proposed mechanism reduces the storage cost $29$S$_\text{SC}$, $34$S$_\text{SC}$ and $39$S$_\text{SC}$ units, respectively. However, additional space required for incorporating the HBAT is $22$S$_\text{SC}$ unit only. Therefore, storage overhead for incorporating the \textit{honeywords} can be compensated from the saved memory. In a nutshell, the proposed questionnaire based authentication technique saves atleast $8$S$_\text{SC}$ unit of storage space (for $q =6$) even after incorporating the HBAT. Thus we may claim that compared to the existing questionnaire based authentication methods, proposed method here incurs no additional storage cost for maintaining the \textit{sweetwords}.

\section{Usability analysis}
\label{u-ana}
Usability analysis in this section has not been confined within the HBAT usability parameters only. We also perform an experimental analyse for capturing users' performance during the login. The users' performances for both the naive login and faster login strategies have been evaluated in terms of login time and error rate. Before going into the experimental analysis, we show how the proposed model fits with HBAT usability parameters.\\

\textbf{System interference and stress on memorability:} Proposed method here neither interferes in the password choice nor imposes any stress on users' mind by forcing the users to remember any additional information. In fact, presented scheme here encourages users to recognize (which is easier than recall) the facts related to the episodic events of their lives. As discussed earlier, recognizing the facts from the episodic memory is an easy task to perform. In this way, proposed questionnaire based authentication system provides adequate user friendliness by fulfilling these two criteria. \\

\textbf{Typo-safety:} For minimizing the typo rates, lead to the false detection, we encourage to generate the option sequences (or \textit{sweetwords}) which are (structure wise) substantially different from each other. To achieve this, we first define the following term.\\

\textbf{Definition 1:} \textbf{\textit{$\lambda-$different strings:}}\textit{Two strings of the same length are said to be $\lambda-$different, if at least in $\lambda$ occasions, two characters belonging to the same index position are different from each other.} \\

Notably, if two strings are $\lambda-$different, then they will also be $(\lambda-1)-$different, provided $(\lambda-1) \geq 0$. However, they will not be $(\lambda+1)-$different.\\

For example, \textit{AAABBA} and \textit{AAADDA} are $2-$different strings as alphabets belonging to $4^{th}$ and $5^{th}$ index positions are different. Also, these two strings are $1-$different, but not $3-$different.\\

\textbf{\textit{Proposition:}} \textit{If the elements of the strings belong to a set of cardinality $\mathcal{C}$, then for two $\lambda-$different strings, s$_1$ and s$_2$, maximum chances of typing s$_2$ mistakenly while entering s$_1$ (or vice versa) can be derived as}
\begin{equation}
\text{P}_{\text{Typing mistake}} = \big(\dfrac{1}{\mathcal{C}-1}\big)^{\lambda}
\label{eqn-4}
\end{equation}

Based on the proposition, we state that for achieving typo-safety, the system should generate the \textit{sweetwords} in such a manner so that all the \textit{sweetwords} must be (pairwise) $\lambda-$different from each other. To meet this objective, we have designed the Algorithm \ref{algo-2} based on the following functions.\\ 

\begin{itemize}\compresslist
\item gStr($\ell$): Returns an arbitrary string (or option sequence) of length $\ell$, constructed by using the elements of the set $\{$A, B, C, D$\}$.
\item add(str, Lst): Appends the string str to the list Lst.
\item comp(str1, str2): For two strings str1 and str2, this function returns an integer value associated with number of times, two alphabets belonging to the same index position do not get matched. For example, comp(AABBCD, ABDACC) will return $4$.
\end{itemize}

\begin{algorithm}[H]
\SetAlgoLined
 initialization: \textit{no$\_$of$\_$sweetwords} $\leftarrow$ $1$, add(act$\_$ops, List);\//* act$\_$ops denotes the actual option sequence*/\\ 
 \While{(no$\_$of$\_$sweetwords $!=k$)}{
  \textit{valid} $\leftarrow 1$\; f$\_$ops $\leftarrow$ gStr($\ell$)\;
  \For{($i =1$ to sizeof(List))}{
   $\eta \leftarrow$ comp(List[i], f$\_$ops)\;  
   \If{($\eta < \lambda$)}{
   \textit{valid} $\leftarrow 0$; \textbf{break}\;
   }
  }
  
  \eIf{(\textit{valid} $:= 1$)}{
  add(f$\_$ops, List)\; \textit{no$\_$of$\_$sweetwords}++\;
  }{\textbf{continue}\;}

 }
 \caption{\textbf{Generating$\_$Sweetwords}($k$, $\lambda$, act$\_$ops, $\ell$)}
 \label{algo-2}
\end{algorithm}

.\\

For $\mathcal{C}$ as $4$ and $q$ as $6, 7$ and $8$, we choose $\lambda$'s value as $3, 4$ and $5$, respectively. Therefore, with reference to these values and Equation \ref{eqn-4}, we may state that proposed method here $96.3\%$, $98.77\%$ and $99\%$  typo safe for $q$'s value as $6$, $7$ and $8$ respectively.  \\

\textbf{Experimental analysis:} For analyzing users' behaviour during login, we took help from $80$ participants ($23$ female and $57$ male). The participants were having correct-to-normal eyesight and were capable of operating computers. The participants were selected from an age group between $19$ and $67$. The login behaviour of the participants under both the \textit{naive} and \textit{faster} login strategies was recorded for the analysis purpose. We varied the value of $q$ from $6$ to $8$ for both these cases ($q'$s value was selected based on the users' performance shown in Table \ref{mem-initial}). It is important to note that according to the report in \citep{no-of-user-guide-1}, most of the studies in this area used a test group between $11$ and $25$, while very few used a test group of $50$ or more participants \citep{no-of-user-guide-2}. Therefore, we believe that a set of $80$ participants provides a better insight about the proposed scheme.\\

At the beginning, we presented the basic motivation behind our work to the participants. We uploaded our scheme to the participants' laptops for training purpose. For each value of $q$, then we followed the steps, mentioned below.

\begin{itemize}\compresslist
\item \textbf{Demonstration of Naive strategy:} We first showed the registration process and the login process using the naive approach.

\item \textbf{Practice and use:} Participants were asked to register into our system. We then gave them time for training purpose. This intention behind this was to allow the participants to become habituated with our scheme.

\item \textbf{Data collection I:} After $2$ days, we collected the test data in our laboratory for capturing the login time and error rate. We denote the login time and error rate obtained in this phase as the initial login time error rate, respectively. 

\item \textbf{More practice and use:} We gave the participants another $5$ days for training purpose for improving their performance.

\item \textbf{Data collection II:} After $5$ days, we again collected the test data. We denote the login time and error rate obtained in this phase as the posterior login time and error rate, respectively.

\item \textbf{Demonstration of Faster login strategy:} Thereafter, for shifting to the next phase of our experimental study, we demonstrated the faster login strategy to all the participants. We observed that most of them found this really interesting as they sensed of achieving improved login time by following this strategy.

\item \textbf{Memorization time:} We gave participants time for memorizing their option sequences for faster login and requested them to come back (earliest possible) for login again. We strongly recommend them to interact more with our scheme for memorizing the option sequences. As mentioned in Section \ref{prop-3}, let the required time for memorizing the option sequence be denoted as $t_\text{mem}$. We measured $t_\text{mem}$ on the basis of login attempts.


\item \textbf{Data collection III:} After $t_\text{mem}$ time, on the basis of their return, we asked each participant for making two attempts for login. We collected the test data again based on the participants' performance in our laboratory. The login time and error rate, in the first and second attempt have been identified as initial and posterior login time and error rate, respectively.
\end{itemize}   

\begin{figure}[!ht]
\centering
\includegraphics[width=0.7\textwidth]{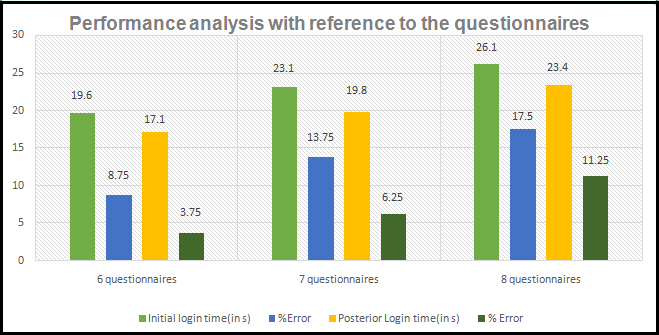}
\caption{Participants' performance based on the collected test data in Data Collection I and II phases.} 
\label{per-1}
\end{figure}

\begin{figure}[!ht]
\centering
\includegraphics[width=0.7\textwidth]{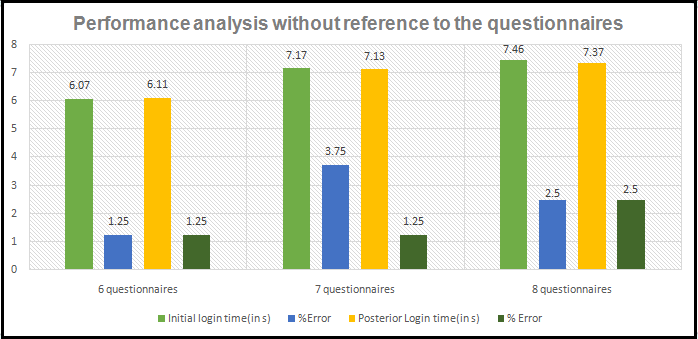}
\caption{Participants' performance based on the collected test data in Data collection III phase.} 
\label{per-2}
\end{figure}

We found that $t_\text{mem}$ for the participants was obtained no more than $15$ login attempts (or $13$ days). Therefore, the experimental study shows that without putting any significant effort, a user can memorize her option sequence. Figure \ref{per-1} and Figure \ref{per-2} illustrate the test results. Moreover, participants' performance in  Figure \ref{per-2} suggests that with a small amount of memorization time, login time can be reduced to a great extent.  

After conducting the experiment, we asked the participants to fill up a feedback form to rate our proposed model. The obtained feedback result was found promising and is presented in Table \ref{feedback}. \\

\begin{table}[!ht]
\centering
\begin{tabular}{|c |c |c|}
\hline
Choices & Agreed participants & Percentage ($\%$)\\
\hline
Love to use & $47$ & $58.75$ \\
\hline
Easy to use & $11$ & $13.75$\\
\hline
Find usable & $15$ & $18.75$\\
\hline
Bit difficult & $5$ & $6.25$\\
\hline
Extremely difficult & $0$ & $0$\\
\hline 
Not sure & $2$ & $2.5$\\
\hline
\end{tabular}
\caption{Recorded feedbacks from the $80$ participants}
\label{feedback}
\end{table}

\textbf{Note 5:} Discussion in this section shows that the proposed method well satisfies the important HBAT usability features along with providing very high typo-safety above $96.3\%$. Also, the experimental study refers that after a short time period of $13$ days, at max, the users can easily remember their option sequence (for all the variations of $q$), and this helps them to finish the login within $8$ seconds (with an error rate no greater than $2.5\%$). In a nutshell, the overall usability analysis shows that after a small training period, proposed system can be used like a conventional password based system to meet all the essential usability criteria of an ideal HBAT.

\section{Security analysis}
\label{s-ana}
In this section, we show how the proposed scheme meets the essential HBAT security criteria, DoS resiliency and flatness.\\

\textbf{Security against the DoS attack:} For each questionnaire having \textit{d} alternatives, an attacker can think of $d^q$ probable option sequences for selected $q$ questionnaires by a user. Therefore, the value of $n$ (see criterion $1$ in Section \ref{dofl}) can be represented as $d^q$ here. For the default value of $d =4$ and $k =20$, the success probability of mounting DoS attack yields to $0.004$, $0.001$ and $0.0002$ for the values of $q$ as $6$, $7$ and $8$, respectively. Therefore, proposed scheme provides robust security against the DoS attack. \\

\textbf{Achieving flatness:} Figure \ref{overview} at the beginning of Section \ref{prop} presents the essential properties that a questionnaire based authentication scheme must satisfy for achieving flatness. Later, in Section \ref{prop-1} and Section \ref{prop-2}, we have shown that proposed scheme holds all these properties and thus generates a flat list of \textit{sweetwords}.

\section{Comparative analysis}
\label{comparative-ana}
In our comparative study, we consider two recent studies in \citep{honeyword-nilesh} and \citep{flatness} as they overcome the limitations of all the previously proposed \textit{honeyword} generation approaches. As mentioned earlier, being a legacy-UI based HBAT, proposed method by \textit{Imran Erguler} tries to meet the essential criteria of an ideal HBAT \citep{flatness}. In contrast, in \citep{honeyword-nilesh}, authors make an effort to meet those same essential criteria under the light of their proposed modified-UI based HBAT. We name the proposed methods in \citep{flatness} and \citep{honeyword-nilesh} as \textit{EH} and \textit{NH}, respectively. \\

Previously, we have shown that proposed HABT here explores episodic memory of the users for imposing no stress on users' mind. Also, proposed scheme does not influence the answer of a user for any question. Thus, as suggested in \citep{honeyword-juels}, proposed method falls under the category of legacy-UI HBAT. 
Based on the presented text so far, Table \ref{comp-ana-01} and \ref{comp-ana-02} compare our scheme with the EH and NH.\\

\begin{table}[!ht]
\centering
\begin{tabular}{|c |c |c |c |c |c|}
\hline
Method & UI type       & Stress-on-mind    & System interference    & Typo-safe & Storage-overhead\\  
\hline
EH     & Legacy        &  No			   &      No				&    Yes	  &  Yes \\
\hline			 	
NH     & Modified      &  Yes		       &      Yes               &    Yes      &  No \\	
\hline
Proposed & Legacy      & No                &      No                &    Yes      &  No $\otimes$ \\ 
\hline 
\end{tabular}
\caption{Above table shows comparative view among the methods from usability and storage cost. The symbol $\otimes$ indicates ``as the additional memory required for storing the \textit{honeywords} is compensated from the saved memory".}
\label{comp-ana-01}
\end{table}   

\begin{table}[!ht]
\centering
\resizebox{0.9\textwidth}{!}{
\begin{tabular}{|c| c| c| c| c| c| c|}
\hline
Method & DoS resiliency & Tg resilient & Ch resilient & Rp proof & Generate-flat-alternatives & Flat \\
\hline
EH	   &  	Weak        &   No         &   No         &   No          &     N/A                    & No\\
\hline
NH     &    Strong      &   No         &   Yes        &   Yes         &     N/A                    & Yes $\otimes$\\
\hline
Proposed &  Strong      &   Yes        &   Yes        &   Yes         &     Yes                    & Yes\\
\hline   
\end{tabular}}
\caption{Above table shows comparative view among the methods from security point of view. Tg and Ch represent targeted guessing and correlational hazardous, respectively. Rp indicates the recognizable pattern. The symbol $\otimes$ refers ``conditionally flat only if risk associated with targeted guessing is not considered".}
\label{comp-ana-02}
\end{table}

Comparative study shows that proposed scheme only meets flatness criterion unconditionally. In contrast, the existing solutions, which are known to provide the best security, cannot even satisfy all criteria for generating the flat \textit{honeywords}. Previously, in Section \ref{erguler-approach}, we have shown incapability of EH in providing adequate security against the DoS attack and hence, Erguler's approach in \citep{flatness} can be categorized as a weak model in handling the DoS. In \citep{honeyword-nilesh}, authors show that NH defends DoS attack with the very high probability $0.91$ and hence NH can be considered as a much stronger security model to address this threat. Following the same path, Section \ref{u-ana} shows that our proposed model always resists DoS attack with higher probability than $0.99$ and ensures very high security against the DoS attack. 

Table \ref{comp-ana-01} shows that proposed method here not only performs well in satisfying all the security parameters but stands strong from the usability perspective too. In addition to this, it is very important to note that storing the alternatives by utilizing least possible storage compensates the overheads for incorporating the \textit{honeywords}. In fact, in Section \ref{s-ana} we show that adopted storage strategy nullifies the cumulative memory overheads.  Therefore, the discussion in this section reveals that proposed method stands tall compared to the existing advanced HBAT.\\

\textbf{Discussion:} In this work, we identify a set of $10$ generic questions belonging to the shareable secret life of the users. The alternatives for $b\_$\textit{questions} have been selected from a group of almost equally probable elements and hence we claim that proposed method generates much improved flatter list of \textit{sweetwords}. In addition to this, all the attributes depicted in Figure \ref{overview} are satisfied to reinforce the flatness standard which is missing in the existing state of arts. From the usability perspective, proposed method does not influence the users' answers or force them to remember any additional information. Moreover, usability standard gets hiked as the users need to recognize (which is much easier than recall) the correct answer based on their episodic memory. Thus, being legacy-UI, proposed approach meet the flatness criterion.   

\section{Conclusion}
\label{conclusion}

Modern research strongly depends on \textit{honeywords} to detect the breach at the server side. Though some significant efforts have been made, existing \textit{honeyword} generation techniques never address all the issues affecting flatness. In this study, we have shown that proposed technique overcomes all the limitations and achieves much improved flatness. In addition to this, our proposal resists the DoS attack with the probability more than $0.99$ to meet all the security parameters.  The proposed scheme gets benefited from the usability perspective as the passwords are formed based on the episodic memory of the users. Also, a high typo-safety has been assured as a typing mistake sets off a false alarm with lesser probability than $0.03$. Along with satisfying both the security and usability parameters, our method provides an efficient way for storing the alternatives, which greatly helps in nullifying the cumulative storage overhead for incorporating the \textit{honeywords}. In future, we will try to identify more generic questions from the shareable secret life of the users and form more uniformly distributed groups for selecting the alternatives. Nonetheless, we believe that this paper shows an efficient approach for generating the \textit{honeywords} which will further benefit the security community from their use.

\section*{References}

\end{document}